\documentclass[pra, twocolumn, showpacs, floatfix, superscriptaddress]{revtex4}
\usepackage{amsmath}
\usepackage{amssymb}
\usepackage{latexsym}
\usepackage{bbm}
\usepackage{amsthm}
\usepackage{graphicx}
\usepackage{times}
\theoremstyle{plain}
\newtheorem{de}{Definition}

\begin{document}
\title{Nonclassical correlation in a multipartite
quantum system: two measures and evaluation}
\author{Akira SaiToh\footnote{The present address is $2$.}}
\email{saitoh@qc.ee.es.osaka-u.ac.jp}
\affiliation{Department of Systems Innovation, Graduate School of Engineering
Science, Osaka University, Toyonaka, Osaka 560-8531, Japan}
\author{Robabeh Rahimi}
\email{rahimi@alice.math.kindai.ac.jp}
\affiliation{Interdisciplinary Graduate School of Science and
Engineering, Kinki University, Higashi-Osaka, Osaka 577-8502, Japan}
\author{Mikio Nakahara}
\affiliation{Department of Physics, Kinki University, Higashi-Osaka,
Osaka 577-8502, Japan}
%\email{nakahara@math.kindai.ac.jp}

\begin{abstract}
There is a commonly recognized paradigm in which a multipartite quantum
system described by a density matrix having no product eigenbasis
is considered to possess nonclassical correlation.
Supporting this paradigm, we define two entropic measures of
nonclassical correlation of a multipartite quantum system.
One is defined as the minimum uncertainty about a joint system after we
collect outcomes of particular local measurements.
The other is defined by taking the maximum over all local systems about
the minimum distance between a genuine set and a mimic set of
eigenvalues of a reduced density matrix of a local system. The latter
measure is based on an artificial game to create mimic eigenvalues of a
reduced density matrix of a local system from eigenvalues of a density
matrix of a global system. Numerical computation of these measures for
several examples is performed.
\end{abstract}
%PACS
%03.65.Ud 	Entanglement and quantum nonlocality
%03.67.-a 	Quantum information
%05.30.Ch 	Quantum ensemble theory
%02.60.-x       Numerical methods (mathematics)

\maketitle
\section{Introduction}
There has been a longstanding discussion on the definition of quantumness
in a quantum state of a multipartite system. One definition is, of course,
entanglement which is considered equivalent to inseparability according to
the separability paradigm \cite{W89, P96-1, P96-2}.
The separability paradigm suggests, as is
well-known in this field, a classification of density matrices of a
system consisting of subsystems $1,\ldots,m$.
Separable density matrices are those of the form
\begin{equation}
 \rho_\mathrm{sep}^{[1,\ldots,m]}=\sum_k w_k \rho_k^{[1]}\otimes
\cdots\otimes\rho_k^{[m]}
\end{equation}
with positive weights $w_k$ ($\sum_kw_k=1$) and density matrices
$\rho_k^{[\cdot]}$ of subsystems. Inseparable density matrices
are those that cannot be represented in this form.
A system (consisting of remote subsystems) represented by a separable
density matrix is regarded as a classically correlated system because
local operations and classical communications
(LOCC; see, {\em e.g.}, \cite{PV06}) can create it from scratch: It can
be prepared remotely when distant persons (Alice, Bob,$\ldots$, Marry)
receive instructions from a common source (Clare).

A bipartite system is a typical system to investigate.
Bipartite separable density matrices are those of the form
\begin{equation}
 \rho_\mathrm{sep}^{[A,B]}=\sum_k w_k \rho_k^{[A]}\otimes\rho_k^{[B]}
\end{equation}
with positive weights $w_k$ ($\sum_kw_k=1$). Bipartite
inseparable density matrices are those that cannot be represented in
this form. One supporting evidence for the paradigm is that, 
bipartite system represented by a separable density matrix does not
violate Bell's inequality \cite{P96-1, P96-2}.

Detection methods of inseparability have
opened a large research field, many of which are based on the Peres-Horodecki
test \cite{P96-1,H96} using positive but not completely
positive linear maps.

It is still a challenging issue to find classes of density matrices
possessing nonlocal nature, other than the class of inseparable density
matrices.
Bennett {\em et al.} \cite{B99} discussed a certain nonlocality about locally
nonmeasurable separable states. Ollivier and Zurek \cite{Z02} later
introduced a measure called quantum discord defined as a discrepancy of
two expressions of a mutual information that should be
equivalent to each other in a classical information theory. Another
branch of study on quantumness was started by Oppenheim and the
Horodecki family \cite{O02}; this was extensively studied by a group
consisting of the Horodecki family and other authors \cite{HHH05}.
They introduced a protocol called closed LOCC (CLOCC).
This protocol allows only local unitary operations, attaching ancillas
in separable pure states, and operations
to send subsystems through a complete dephasing channel. They also
defined a measure of quantumness named quantum deficit as a discrepancy
between the information that can be localized by applying CLOCC operations
and the total information of the system. The present work is in the
stream of these studies on quantumness in correlation.

One way to evoke a discussion on the validity of the separability
paradigm is to look at the persistent question why a pseudo-pure state 
\begin{equation}\label{eqpseu}
\rho_\mathrm{ps}=p|\psi\rangle\langle \psi|+(1-p)\mathbbm{1}/d
\end{equation}
with $|\psi\rangle$ an entangled pure state and $d$ the dimension of
the Hilbert space, is often regarded as a classically correlated state
for small probability $p$ because of its separability proved by
Braunstein {\em et al.} \cite{B99-2}. The state $\rho_\mathrm{ps}$ can
be regarded as a state possessing quantumness in correlation if we
choose another paradigm than the separability paradigm. It is thus a
rather conceptual question and is in relation to the following
discussion.

Suppose that we have a system consisting
of two subsystems (local systems) and cannot eliminate a local and/or global
superposition by local unitary operations. A system described by
$\rho_\mathrm{ps}$ for any $p\not =0$ is a typical example. Then, can
correlation between those local systems be
regarded as classical one? An answer is found in the paper by
Oppenheim {\em et al.} \cite{O02} (see also \cite{HHH05}) in which
they made use of a class of states having a biproduct eigenbasis for a
certain classical/nonclassical separation.
\begin{de}
Let us consider a joint system consisting of subsystems $A$ and $B$
with Hilbert space dimensions $d^{[A]}$ and $d^{[B]}$, respectively.
A complete orthonormal basis (CONB) consisting of eigenvectors,
$\{|e_i^{[A,B]}\rangle\}_{i=1}^{d^{[A]} d^{[B]}}$, 
of a density matrix of the system is a biproduct eigenbasis iff it is
given by the direct product as
$\{|e_i^{[A,B]}\rangle\}_i = \{|e_j^{[A]}\rangle\}_{j=1}^{d^{[A]}}
\times\{|e_k^{[B]}\rangle\}_{k=1}^{d^{[B]}}$
where $\{|e_j^{[A]}\rangle\}_j$ and $\{|e_k^{[B]}\rangle\}_k$ are eigenbases of
individual subsystems.
\end{de}
If a density matrix $\rho^{[A,B]}$ has a biproduct (BP) eigenbasis
$\{|e_i^{[A,B]}\rangle\}_i = \{|e_j^{[A]}\rangle\}_{j=1}^{d^{[A]}}
\times\{|e_k^{[B]}\rangle\}_{k=1}^{d^{[B]}}$,
it can be written in the form
\begin{equation}\label{formBP}
 \rho^{[A,B]}_\mathrm{BP} = \sum_{jk} c_{jk}|e_j^{[A]}\rangle\langle e_j^{[A]}|
\otimes |e_k^{[B]}\rangle\langle e_k^{[B]}|
\end{equation}
with coefficients $0\le c_{jk}\le 1$ ($\sum_{jk}c_{jk}=1$). A state
represented by this density matrix is called a properly classically
correlated state or, shortly, a classically correlated state \cite{HHH05}.
It is also called classical-classical state \cite{PHH08}.
The state that cannot be represented by a density matrix in the above
form is called a nonclassically correlated state.

\begin{figure}[pbt]
\begin{center}
\scalebox{0.65}{\includegraphics{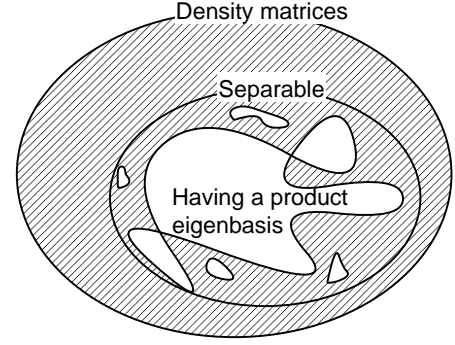}}
\caption{\label{figNCC1}
Class of density matrices having a product eigenbasis (white regions in
the class of separable density matrices) and the
class of density matrices having no product eigenbasis (shaded region).
}
\end{center}
\end{figure}

It is trivial to extend the above definition to a multipartite system.
A density matrix having a (fully) product (FP) eigenbasis is written in
the form:
\begin{equation}\label{formFP}
 \rho^{[1,\ldots,m]}_\mathrm{FP}=
\sum_{j,\ldots,x=1,\ldots,1}^{d^{[1]},\ldots,d^{[m]}}
c_{j,\ldots,x} |e_j^{[1]}\rangle\cdots |e_x^{[m]}\rangle
\langle e_j^{[1]}|\cdots \langle e_x^{[m]}|,
\end{equation}
with the local CONBs $\{|e_j^{[1]}\rangle\}_{j=1}^{d^{[1]}}, \ldots,
\{|e_x^{[m]}\rangle\}_{x=1}^{d^{[m]}}$ ($d^{[\cdot]}$ is the dimension of
the Hilbert space of a local system) and the coefficients 
$0\le c_{j,\ldots,x}\le 1$ ($\sum_{j,\ldots,x}c_{j,\ldots,x}=1$) in use.
This definition separates the class of density matrices having a product
eigenbasis (this is a non-convex set) from the class of density matrices
having no product eigenbasis, as illustrated in Fig.\ \ref{figNCC1}. The
latter class is characterized by non-vanishing superposition (namely,
non-vanishing off-diagonal elements) under local unitary transformations.

In relation to the discussion on biproduct eigenbasis, Groisman {\em et
al.} \cite{GKM07} recently introduced a measure of quantumness
given by
\[
\mathcal{Q}(\rho^{[A,B]})=\underset{\rho_\mathrm{BP}^{[A,B]}}{\mathrm{min}}
F(\rho^{[A,B]},\rho_\mathrm{BP}^{[A,B]}),
\]
where the minimum is taken over all density matrices
$\rho_\mathrm{BP}^{[A,B]}$ having a biproduct eigenbasis; $F$ is any
properly-defined distance function, such as the
relative entropy function. $\mathcal{Q}(\rho^{[A,B]})$ must be invariant
under local unitary operations and must be zero for $\rho^{[A,B]}$ having a
biproduct eigenbasis. They also suggested to use a special density
matrix $\rho_\mathrm{Sch}^{[A,B]}$ (they called it Schmidt state) to
define an easily-computable measure of quantumness
$F(\rho^{[A,B]},\rho_\mathrm{Sch}^{[A,B]})$. The density matrix
$\rho_\mathrm{Sch}^{[A,B]}$ is created by keeping only diagonal
elements of $\rho^{[A,B]}$ under the special basis diagonalizing 
$\mathrm{Tr}_B\rho^{[A,B]}\otimes\mathrm{Tr}_A\rho^{[A,B]}$.

Indeed, it is a natural statement that a measure $M$ of
nonclassical correlation should satisfy the conditions:\\
(i) $M=0$ if a system is described by a density matrix having a product
eigenbasis (i.e., $M=0$ is a necessary but not sufficient condition for
a state to have a product eigenbasis).\\
(ii) $M$ is invariant under local unitary operations.\\
These conditions are considered to be prerequisite hereafter.
In addition, one may test if a measure possesses either of (iii) full
additivity, (iv) weak additivity, (v) subadditivity, etc. in the family
of additivity properties. These properties are defined in the following
way. They are desirable properties for measures of multipartite
correlation and not exactly based on the additivity concept for
bipartite correlation often seen for entanglement measures \cite{HHH00}.
Let us denote a measure of $m$-partite nonclassical correlation by
$M_m(\sigma)$ where $\sigma$ is the density matrix of an $m$-partite
quantum system. First, the measure is fully additive if and only if 
\[
M_{m_1\times m_2}(\sigma_1\otimes\sigma_2)=
M_{m_1}(\sigma_1)+M_{m_2}(\sigma_2)
\]
with $\sigma_1$ the density matrix of an $m_1$-partite system and
$\sigma_2$ the density matrix of an $m_2$-partite system.
Second, the measure possesses weak additivity if and only if
\[
M_{m^n}(\sigma^{\otimes n})=nM_{m}(\sigma).
\]
Third, the measure possesses subadditivity if and only if
\[
M_{m_1\times m_2}(\sigma_1\otimes\sigma_2)\le
M_{m_1}(\sigma_1)+M_{m_2}(\sigma_2).
\]

In this paper, we introduce two measures of nonclassical correlation for
a general multipartite system and numerically evaluate them for several
examples. One of them, defined in Sec.\ \ref{sec2}, is similar to but
different from the measure proposed by Groisman {\em et al.} and the
other one, defined in Sec.\ \ref{sec3}, is totally independent.
This paper is organized as follows.
In Sec.\ \ref{sec2}, we quantify a nonclassical correlation by a
measure defined as the minimum uncertainty with respect to a joint
system after we collect outcomes of particular local measurements.
This measure satisfies the full additivity condition. The other measure
will be introduced in Sec.\ \ref{sec3}, which is defined in the
following way: we consider the minimum distance between a genuine set
and a mimic set of eigenvalues of a reduced density matrix of a local
system on the basis of an artificial game in which one creates
mimic eigenvalues of a reduced density matrix of a local system from
eigenvalues of a density matrix of a global system. The measure is
defined by taking the maximum of this minimum distance over all local
systems. It satisfies the subadditivity condition and a slightly
stronger condition. We perform numerical computation of the
two introduced measures for several examples in Sec.\ \ref{sec4} and
compare them with negativity that is a common entanglement measure
based on the separability paradigm. A discussion on definitions of
nonclassical correlation is given in Sec.\ \ref{sec5}. Section
\ref{sec6} summarizes the results of this paper.

\section{Measure of nonclassical correlation I}\label{sec2}
We introduce the first of two measures of nonclassical correlation in this
section. It is based on the paradigm in which a system described by a
density matrix having a product eigenbasis is considered to possess
only a classical correlation; in contrast, a system described by a density
matrix having no product eigenbasis is considered to possess a
nonclassical correlation. We have seen the form of a density matrix
having a product eigenbasis for a bipartite system in Eq.\ (\ref{formBP})
and that for a multipartite system in Eq.\ (\ref{formFP}).
We will begin with the bipartite case.
\subsection{Bipartite case}
\begin{figure}[pbt]
\begin{center}
\scalebox{0.4}{\includegraphics{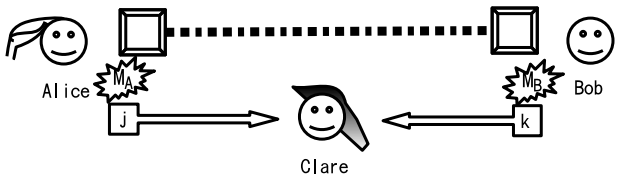}}\caption{\label{figABC}
System for which nonclassical correlation between Alice's part and
Bob's part of a quantum state is discussed. A measure of
nonclassical correlation is defined as a minimum uncertainty for Clare
about the state after receiving their reports.}
\end{center}
\end{figure}

To quantify nonclassical correlation between distant subsystems of a
bipartite system, we consider the situation illustrated in
Fig.\ \ref{figABC}.
Let us introduce Alice, Bob, and Clare. Alice and Bob have subsystems
of a system and they are distant from each other. They can send reports to
Clare. Alice/Bob can choose a complete orthonormal basis of her/his
subsystem (basis $\{|e_j^{[A]}\rangle\}_j$ for Alice and basis
$\{|e_k^{[B]}\rangle\}_k$ for Bob) for local (projective) measurements.
Suppose that Alice and Bob use the observables
$M_A=\sum_j j |e_j^{[A]}\rangle\langle e_j^{[A]}|$
and $M_B=\sum_k k |e_k^{[B]}\rangle\langle e_k^{[B]}|$,
respectively, and report the outcomes $j$ and $k$ to Clare.
The probability that Clare receives $j$ from Alice and $k$ from Bob is
$p_{jk}=
\langle e_j^{[A]}|\langle
e_k^{[B]}|\rho^{[A,B]}|e_j^{[A]}\rangle|e_k^{[B]}\rangle$.
The same process is performed for many copies of the same system shared
by Alice and Bob without changing the bases initially chosen.
Then, minimum uncertainty (over all possible initial choices of local bases)
that Clare has about $\rho^{[A,B]}$ after receiving their reports is
\begin{equation}\label{eqD}
D(\rho^{[A,B]})=
\underset{\mathrm{local~bases}}{\mathrm{min}}
\left(-\sum_{jk}p_{jk}\log_2 p_{jk}\right) - S_\mathrm{vN}(\rho^{[A,B]}),
\end{equation}
where $S_\mathrm{vN}(\rho^{[A,B]})$ is the von Neumann entropy.
We employ this quantity as a measure of nonclassical correlation.
It is obvious that $D(\rho^{[A,B]}) = 0$ for a density matrix with a
biproduct eigenbasis. Otherwise, there is a possibility that
$D(\rho^{[A,B]}) > 0$. Thus a bipartite separable density matrix having no
product eigenbasis (as well as a bipartite inseparable density matrix)
possibly has a nonlocal correlation that may be quantified by $D(\rho^{[A,B]})$.
A typical example is a density matrix shown in Eq.\ (\ref{eqpseu}) for
$0<p\le1$. For this density matrix, $D(\rho^{[A,B]})>0$ holds.
In addition, we should stress that
$D(\rho^{[A,B]})$ is invariant under local unitary operations; this is
clear from its definition in which we search over all local bases to
obtain the minimum.

One may find that the above process involving Alice, Bob, and Clare
can be reconstructed in terms of CLOCC operations \cite{O02,HHH05}
to define the same quantity: Consider the minimum discrepancy between
the von Neumann entropy of the original state $\rho^{[A,B]}$ and that of
the state Clare can achieve by merging states received from Alice and
Bob only after Alice and Bob locally use dephasing operations
$\Lambda_D(\rho^{[X]})=\sum_{i=1}^{d^{[X]}}\langle
e_i|\rho^{[X]}|e_i\rangle|e_i\rangle\langle e_i|$ where $X$ is
subsystem $A$ or $B$, $d^{[X]}$ is the dimension of its Hilbert space,
and $\{|e\rangle_i\}_i$ is the CONB of her/his choice. The minimum is
taken over all choices of local CONBs. Then the discrepancy is equal to
the measure $D(\rho^{[A,B]})$. In this way, $D(\rho^{[A,B]})$ can be
related to the CLOCC protocol that was the base protocol for quantum
deficit \cite{HHH05}. We should note that, in general,
$D(\rho^{[A,B]})$ is not equal to quantum deficit; this is clear by
comparing Eq.\ (11) of Ref.\ \cite{HHH05} with the above definition. The
measure $D(\rho^{[A,B]})$ is equal to the quantum deficit in the case
where the zero-way CLOCC protocol \cite{HHH05} is considered \footnote{
The zero-way setting of CLOCC is as follows \cite{HHH05}:
there are two parties allowed to communicate under CLOCC only
after local complete dephasing possibly subsequent to local unitary
operations. In general, the quantum deficit is equal to
${\rm min}_{\Lambda\in{\rm CLOCC}}
[S_{\rm vN}({\rho'}_{\rm Alice}) + S_{\rm vN}({\rho'}_{\rm Bob})]
- S_{\rm vN}(\rho^{[A,B]})$ with $\rho'=\Lambda(\rho^{[A,B]})$.
In case of zero-way CLOCC, the minimum is obtained for the case where
Alice or Bob has $\rho'$ totally and the other person has a null system.
Then the zero-way case quantum deficit is equal to
${\rm min}_{\Lambda\in{\rm zero-way CLOCC}} 
S_{\rm vN}(\rho')- S_{\rm vN}(\rho^{[A,B]})$. This is equal to $D$ for
the bipartite case.}.

In addition, the measure $D(\rho^{[A,B]})$ is similar to but
not included in the measure of quantumness defined by Groisman
{\em et al.} \cite{GKM07} in the sense that we search over all local
bases while their measure is defined by taking a minimum over all
classical states (hence, over all possible combinations of eigenvalues
and local bases).

\subsection{Multipartite case}
It is straightforward to extend the definition Eq.\ (\ref{eqD}) of
$D$ to that for general multipartite systems. Let us consider a
density matrix $\rho^{[1,\ldots,m]}$ of an $m$-partite system.
Consider local CONBs $\{|e_j^{[1]}\rangle\}_j, \ldots,
\{|e_x^{[m]}\rangle\}_x$.
Then, a measure of nonclassical correlation is given by
\begin{equation}\label{eqDM}\begin{split}
D(\rho^{[1,\ldots,m]})=&
\underset{\mathrm{local~bases}}{\mathrm{min}}
\left(-\sum_{j,\ldots, x}p_{j,\ldots, x}\log_2 p_{j,\ldots, x}\right)\\
&-S_\mathrm{vN}(\rho^{[1,\ldots,m]})
\end{split}\end{equation}
with \[
p_{j,\ldots,x}=\langle e_j^{[1]}|\langle e_k^{[2]}|\cdots \langle e_x^{[m]}|
\rho^{[1,\ldots,m]}|e_j^{[1]}\rangle |e_k^{[2]}\rangle\cdots
|e_x^{[m]}\rangle.
\]
An interpretation of this measure from an operational viewpoint is possible.
Let $L_1$ ($L_2$) be the average code length of an optimal classical
data compression, focusing on only diagonal elements, acting on a 
register of qudits subsequent to local (global) unitary operations.
Then the minimum discrepancy between
$L_1$ and $L_2$ leads to the definition of the measure.  
The value of $D(\rho^{[1,\ldots,m]})$ is zero if $\rho^{[1,\ldots,m]}$
has a (fully) product eigenbasis. In addition,
$D(\rho^{[1,\ldots,m]})$ is invariant under local unitary operations as
is clear from its definition in which the minimum is taken over all
local bases. Furthermore, we can prove that it is fully additive.

The proof of full additivity uses only a general property of an entropy
function in a particular form. Suppose that there is a system consisting
of subsystems $X$ and $Y$ described by the reduced density matrices
$\sigma_X$ and $\sigma_Y$, respectively. Consider an entropy function
$E$ of the product density matrix $\sigma_X\otimes\sigma_Y$ and CONBs
$\{|x\rangle\}$, $\{|y\rangle\}$ of the reduced density matrices, written as
\[\begin{split}
 &E(\{|x\rangle\},\{|y\rangle\},\sigma_X\otimes\sigma_Y)\\
&= -\sum_{|x\rangle|y\rangle}\langle x|\langle y|
\sigma_X\otimes \sigma_Y|x\rangle|y\rangle \log_2 \langle x|\langle y|
\sigma_X\otimes \sigma_Y|x\rangle|y\rangle.
\end{split}\]
We rewrite it as
\[
E(\{|x\rangle\},\{|y\rangle\},\sigma_X\otimes\sigma_Y)=
E(\{|x\rangle\},\sigma_X)+E(\{|y\rangle\},\sigma_Y)
\]
with entropies of subsystems
\begin{eqnarray}
&E(\{|x\rangle\},\sigma_X)=-\sum_{|x\rangle}\langle x|\sigma_X|x\rangle
\log_2 \langle x|\sigma_X|x\rangle,\nonumber\\
&E(\{|y\rangle\},\sigma_Y)=-\sum_{|y\rangle}\langle y|\sigma_Y|y\rangle
\log_2 \langle y|\sigma_Y|y\rangle.\nonumber
\end{eqnarray}
It is obvious that $E(\{|x\rangle\},\{|y\rangle\},\sigma_X\otimes\sigma_Y)$
is minimized if and only if $E(\{|x\rangle\},\sigma_X)$ and
$E(\{|y\rangle\},\sigma_Y)$ are individually minimized.
We conclude that $D(\sigma\otimes\tau)=D(\sigma)+D(\tau)$ holds
for density matrices $\sigma$ and $\tau$ accordingly.
Thus the measure $D$ satisfies the full additivity condition.

\subsection{Numerical method to estimate $D$}\label{secNM}
\begin{figure}[pbt]
\scalebox{0.6}{\includegraphics{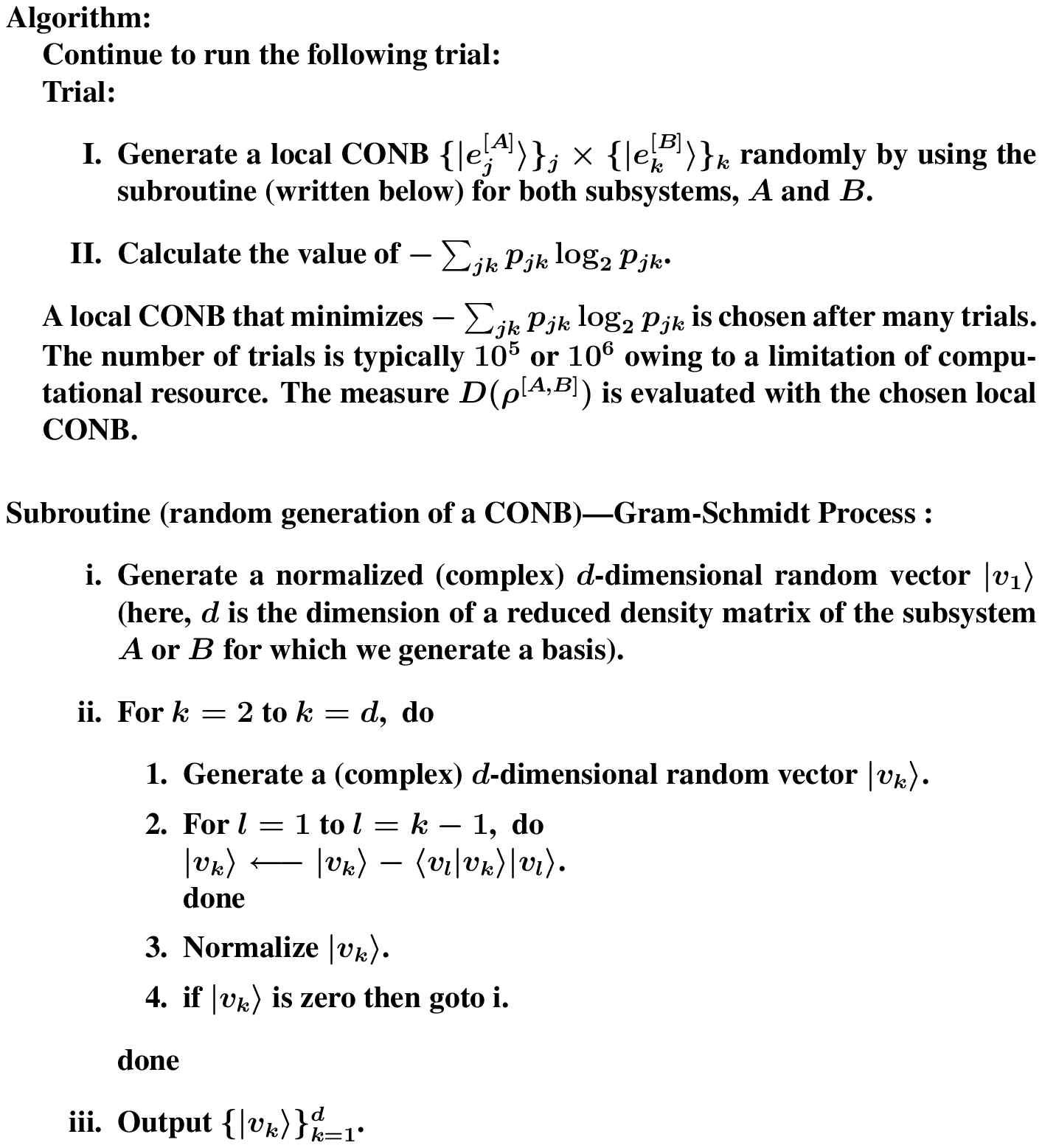}}
\caption{\label{figAlg}
Algorithm to compute an estimated value of $D(\rho^{[A,B]})$.
See the text for the symbols used herein.}
\end{figure}

A pure random search of local bases is a practical method to compute
an estimated value of $D(\rho^{[1,\ldots,m]})$, defined by Eq.\ (\ref{eqDM}),
for a small size of a multipartite density matrix $\rho^{[1,\ldots,m]}$.
Let us consider the bipartite case for clarity, where the density matrix is
$\rho^{[A,B]}$. 
We employ an {\em ad hoc} random search algorithm introduced in
Fig.\ \ref{figAlg} in which a Gram-Schmidt process (see, {\em e.g.}, page
108 of Ref.~\cite{LT85}) is utilized.
In addition, it is needless to say that a similar algorithm can be used
to estimate a value of $D(\rho^{[1,\ldots,m]})$ for a general $m$
numerically.

The random CONB generation is equivalent to a generation of random
unitary matrices acting on fixed CONBs $\{|j\rangle\}$ and
$\{|k\rangle\}$. This is clear from the fact that for any
local unitary operations $U_A$ and $U_B$, we have
$U_A|j\rangle=|e_j\rangle$ and $U_B|k\rangle=|e_k\rangle$.
The probabilities $p_{jk}$ are the diagonal elements
of $U_A^\dagger\otimes U_B^\dagger\rho^{[A,B]}U_A\otimes U_B$.
(It is now trivial to generalize this to the multipartite case.)
Thus random matrix theories \cite{RMT} will be
hopefully used to refine the algorithm. The present algorithm is still
fast enough to estimate a value of $D(\rho^{[1,\ldots,m]})$ for a small
number of qubits in a reasonable time. We show numerical results
in Sec.\ \ref{sec4}.

\section{Measure of nonclassical correlation II}\label{sec3}
The previous measure, $D$, is defined as a minimum uncertainty about
a global system after collecting outcomes of local measurements.
Although the definition itself looks quite reasonable, we have to stress 
the difficulty to find the value of $D$ because we need to try
all possible product eigenbases to find the minimum, in principle. Strictly
speaking, we need to try an infinite number of product eigenbases, but
we rather find an estimate value by using a random search as we have seen.
In this section, we will introduce the second measure derived from an
artificial game to mimic a set of eigenvalues of a reduced density
matrix of a local system by using a set of eigenvalues of a density
matrix of a global system. The advantage of this measure is that it can
be calculated deterministically within a finite time. The disadvantage
is its artificial definition, which may be insignificant taking account of
its advantage. 

\subsection{Bipartite case}
\begin{figure}[pbt]
\begin{center}
\scalebox{0.4}{\includegraphics{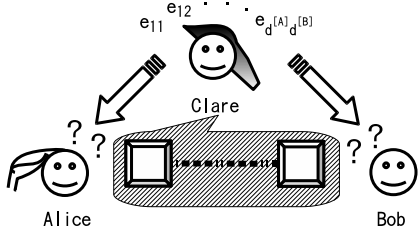}}\caption{\label{figABC2}
Illustration of the artificial game in which Alice (Bob) tries to
find the eigenvalues of the reduced density matrix of her (his) part by
the strategy described in the text. Clare knows the eigenvalues of the
total system.}
\end{center}
\end{figure}

Consider an artificial game pertaining to Alice, Bob, and Clare.
A system consisting of two parts, $A$ (Alice's part) and $B$ (Bob's
part), is described by the density matrix $\rho^{[A,B]}$.
Alice and Bob cannot measure the system at all. Clare knows the set
of all eigenvalues $\{e_{jk}\}_{jk=11}^{d^{[A]}d^{[B]}}$ where $d^{[A]}$
and $d^{[B]}$ are the dimension of the Hilbert space of Alice's part
and that of Bob's part, respectively. Alice (Bob) wants to know the
eigenvalues of her (his) part.  This setup is illustrated in
Fig.\ \ref{figABC2}.
Let us concentrate on Alice's strategy. She asks Clare to
send all the eigenvalues. Alice will partition $d^{[A]}\times d^{[B]}$
eigenvalues received from Clare into $d^{[A]}$ sets 
\[
\{a_{1,1},\cdots,a_{1,d^{[B]}}\},
\cdots, \{a_{d^{[A]},1},\cdots,a_{d^{[A]},d^{[B]}}\}.
\]
 To mimic an eigenvalue of the reduced density
matrix of $A$, she sums up all elements of each set to make a set of
$d^{[A]}$ mimic eigenvalues:
\[
 \{\tilde e_1,\cdots,\tilde e_j,\cdots,\tilde e_{d^{[A]}}\},
\] 
where each element is calculated as
\[
 \tilde e_j = \sum_{k=1}^{d^{[B]}}a_{j,k}.
\]
The number of possible combinations to partition $d^{[A]}\times d^{[B]}$
eigenvalues into $d^{[A]}$ sets is 
$\left(\begin{array}{c}d^{[A]}\times d^{[B]}\\ d^{[B]}\end{array}\right)
\left(\begin{array}{c}(d^{[A]}-1)\times d^{[B]}\\ d^{[B]}\end{array}\right)
\cdots
\left(\begin{array}{c}d^{[B]}\\ d^{[B]}\end{array}\right)$.
Let us write the genuine eigenvalues of the reduced density matrix of
$A$ by $\{e_j\}_{j=1}^{d^{[A]}}$.
The minimum uncertainty for Alice with respect to the set of these
eigenvalues in the artificial game may be given by
\begin{equation}
 F_A(\rho^{[A,B]})=\underset{\mathrm{partitionings}}{\mathrm{min}}\left|
\sum_{j} (\tilde e_j\log_2 \tilde e_j - e_j\log_2 e_j)
\right|.
\end{equation}
We may also consider the minimum uncertainty $F_B(\rho^{[A,B]})$
for Bob with respect to the set of eigenvalues for his local part in the
same game. The larger one of their minimum uncertainties is then given by
\begin{equation}\label{eqG}
G(\rho^{[A,B]})=\mathrm{max}\{F_A(\rho^{[A,B]}), F_B(\rho^{[A,B]})\}.
\end{equation}
This function can be used as a measure of nonclassical correlation
since $G(\rho^{[A,B]})=0$ if $\rho^{[A,B]}$ has a product eigenbasis.
In addition, $G(\rho^{[A,B]})$ is invariant under local unitary
operations according to the definition since local unitary operations
preserve the eigenvalues of reduced density matrices of individual
components and those of the density matrix of the total system.

\subsection{Multipartite case}
An extension of the measure $G$ to a multipartite case is
straightforward. Let us consider an artificial game to find out
eigenvalues of the reduced density matrix of a subpart from eigenvalues
of the density matrix of the total system.
Suppose that Kate has the $k$th part of an $m$-partite
quantum system.  Let the dimension of the Hilbert space of the $k$th
part be $d^{[k]}$ and that of the Hilbert space of the total system
be $d_{\mathrm{tot}}$. She wants to know the eigenvalues
$\{e_j^{[k]}\}_{j=1}^{d^{[k]}}$ of the reduced density matrix of the
$k$th part. Kate receives $d_{\mathrm{tot}}$ eigenvalues from Tony who knows
the eigenvalues of the total system. Kate partitions them into $d^{[k]}$
sets. Summing up elements in individual sets, she has $d^{[k]}$ mimic
eigenvalues $\{\tilde e_j^{[k]}\}$. Thus a measure of nonclassical
correlation for Kate can be
\[
 F_k(\rho^{[1,\ldots,m]})=\underset{{\mathrm{partitionings}}}{{\mathrm{min}}}
\left|\sum_{j=1}^{d^{[k]}}(\tilde e_j^{[k]}\log_2 \tilde e_j^{[k]}
- e_j^{[k]}\log_2 e_j^{[k]})\right|.
\]
We may take the maximum over $k$ to have the measure
\begin{equation}\label{eqGM}
 G(\rho^{[1,\ldots,m]})=\underset{{k}}{{\mathrm{max}}}~F_k(\rho^{[1,\ldots,m]}).
\end{equation}
This is equal to zero if $\rho^{[1,\ldots,m]}$ has a (fully) product
eigenbasis. In addition, it is invariant under local unitary operations
as is clear from the fact that these operations preserve the eigenvalues
of a density matrix of the total system and those of the reduced density
matrices of individual components.
The measure does not satisfy the (full or weak) additivity condition
because the number of possible choices of partitioning eigenvalues grows
rapidly as the system size grows. We can, however, prove its subadditivity.

The proof for its subadditivity is accomplished in the following way.
Suppose the following story: there is a density matrix $\sigma$ of an
$m_\sigma$-partite system having the set of eigenvalues,
$\{\alpha_a\}_{a=1}^{d_\sigma}$ (here, $d_\sigma$ is the
dimension of the Hilbert space of the system).
With this set of eigenvalues, Kate has created a set of
$d^{[k]}$ mimic eigenvalues of the reduced density matrix of the $k$th
part. There is another density matrix $\tau$ of an $m_\tau$-partite system
having the set of the eigenvalues, $\{\beta_b\}_{b=1}^{d_\tau}$.
With this set of eigenvalues, Leo has created a set of
$d^{[l]}$ mimic eigenvalues of the reduced density matrix of the $l$th
part.
Then a new game starts with a joint state $\sigma\otimes\tau$.
The game for Kate/Leo is to make a set of mimic eigenvalues for
her/his part by partitioning the product set of eigenvalues
$\{\alpha_a\}\times\{\beta_b\}$ of $\sigma\otimes\tau$. 
Then, Kate (Leo) may make the same set of mimic eigenvalues as before
because she (he) can make the set
$\{\alpha_a\}$ ($\{\beta_b\}$) firstly by partitioning
$\{\alpha_a\}\times\{\beta_b\}$.
Thus $F_k(\sigma\otimes\tau)\le F_k(\sigma)$ and
 $F_l(\sigma\otimes\tau)\le F_l(\tau)$ hold.
Hence the next inequalities are satisfied.
\[
 G(\sigma\otimes\tau)\le\mathrm{max}\{G(\sigma),G(\tau)\}\le G(\sigma)+G(\tau).
\] 
In this way, the subadditivity $G(\sigma\otimes\tau)\le
G(\sigma)+G(\tau)$ has been proved. We found that a slightly stronger
condition
$G(\sigma\otimes\tau)\le\mathrm{max}\{G(\sigma),G(\tau)\}$ is satisfied
as a result. This property can be named {\it submaximizability}.

\section{Numerical results}\label{sec4}
\subsection{Examples for bipartite cases}
We compare the measures $D$ and $G$ with the negativity \cite{Z98,VW02}
$\mathcal{N}(\rho^{[A,B]})=\frac{\|(I\otimes\Lambda_\mathrm{T})\rho^{[A,B]}\| -
1}{2}$ (here the map $\Lambda_\mathrm{T}$ is the transposition map
acting on $B$).

The first example is the two-qubit pseudo-pure state
\[
\rho_\mathrm{ps}=p|\psi\rangle\langle\psi| + (1-p)\mathbbm{1}/4
\]
with $|\psi\rangle=(|00\rangle+|11\rangle)/\sqrt{2}$.
We used a pure (numerical) random search of local bases, as introduced in
Sec.\ \ref{secNM}, to estimate a value of $D(\rho_\mathrm{ps})$.
The number of trials of local bases is $4.0\times 10^4$
for each data point (namely, for each $p$ in this example) of
$D(\rho_\mathrm{ps})$. (The number of trials is $4.0\times 10^4$ for
the other examples of two-qubit cases and is $4.0\times 10^5$ for the
examples involving an $8\times 8$ density matrix.)
Computation of a value of $G(\rho_\mathrm{ps})$ is, in contrast,
performed analytically. The eigenvalues of $\rho_\mathrm{ps}$ are
$(1+3p)/4$ and $(1-p)/4$ with the multiplicity three for the latter one.
The eigenvalue of the reduced density matrix of a subpart is $1/2$
with the multiplicity two. Thus we have
\[
 G(\rho_\mathrm{ps})=1-H\left(\frac{1+p}{2}\right),
\]
where 
\[
H(x)=-x\log_2x-(1-x)\log_2(1-x)
\]
is the binary entropy function ($0\le x\le 1$).

\begin{figure}[bpt]
\begin{center}
\scalebox{0.7}{\includegraphics{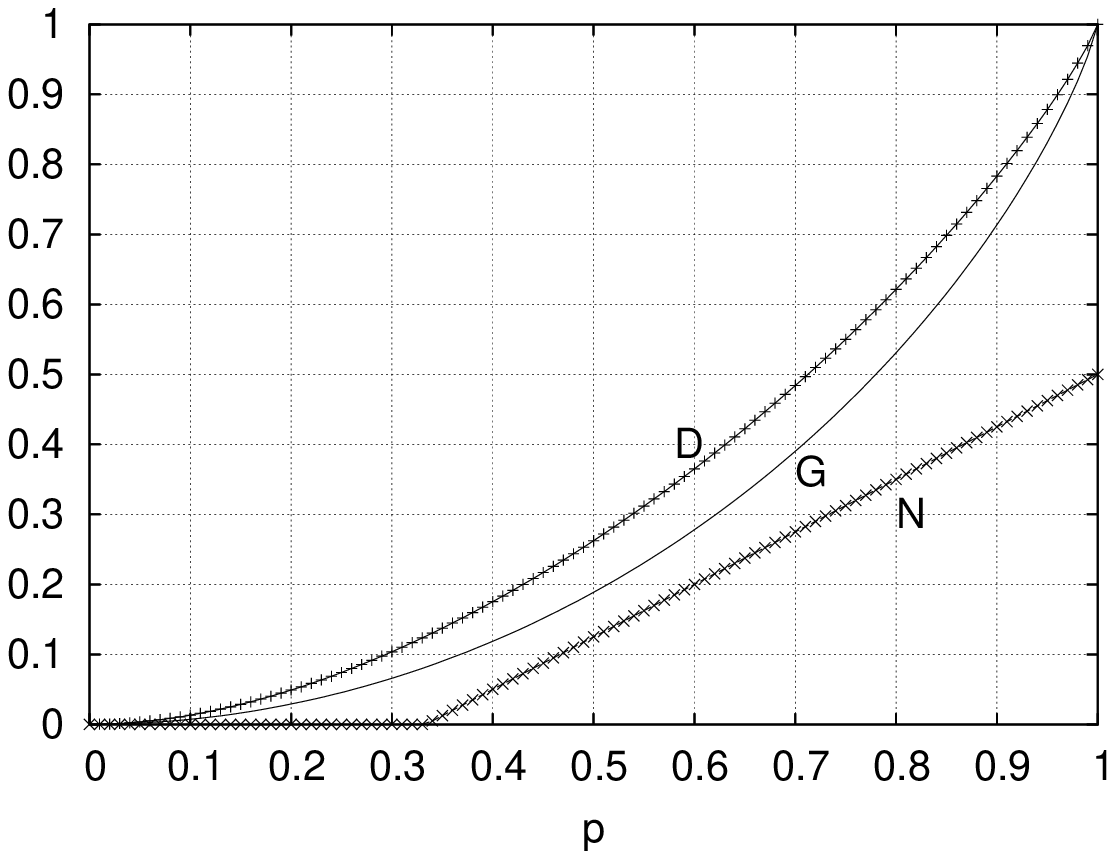}}
\caption{\label{fignume}Plots of $D(\rho_\mathrm{ps})$,
$G(\rho_\mathrm{ps})$, and
$\mathcal{N}(\rho_\mathrm{ps})$ against $p$.} 
\end{center}
\end{figure}

Figure \ref{fignume} shows the plots of $D(\rho_\mathrm{ps})$,
$G(\rho_\mathrm{ps})$, and $\mathcal{N}(\rho_\mathrm{ps})$
against $p$ ($0\le p\le1$).
The measures $D$ and $G$ reflect nonclassical
correlation for $\forall p$ except for $p=0$.

The next example is a mixture of Bell basis states,
represented by the density matrix 
\[
\rho_\mathrm{b}=p|b_1\rangle\langle b_1|+(1-p)|b_2\rangle\langle b_2|
\]
with $|b_1\rangle=(|00\rangle+|11\rangle)/\sqrt{2}$
and $|b_2\rangle=(|01\rangle+|10\rangle)/\sqrt{2}$.
This density matrix has a product eigenbasis if $p=0.5$:
$\rho_\mathrm{b}(p=0.5)=
\frac{1}{2}(|+\rangle|+\rangle\langle +|\langle +|
+|-\rangle|-\rangle\langle -|\langle -|)$ 
with $|\pm\rangle=(|0\rangle\pm|1\rangle)/\sqrt{2}$.

With the same basis search method as the previous example, we can
estimate the values of $D(\rho_\mathrm{b})$.
For this example, however, it can be analytically found: The von
Neumann entropy of $\rho_\mathrm{b}$ is given by $H(p)$ and the
remaining term of Eq.\ (\ref{eqD}) becomes $1$ irrespectively of the choice
of a local basis. Thus
\[
D(\rho_\mathrm{b})=1-H(p).
\]
In addition, analytical computation of
$G(\rho_\mathrm{b})$ is easy: the eigenvalues of $\rho_\mathrm{b}$ are
$p$, $1-p$, and $0$ with the multiplicity two. The eigenvalue of the
reduced density matrix of a subpart is $1/2$ with the multiplicity
two. Thus, taking the minimum over partitionings of the eigenvalues of
$\rho_\mathrm{b}$, we have
\[
 G(\rho_\mathrm{b})=1-H(p).
\]
We find that $D(\rho_\mathrm{b})=G(\rho_\mathrm{b})$ for this particular
example.

\begin{figure}[bpt]
\begin{center}
\scalebox{0.7}{\includegraphics{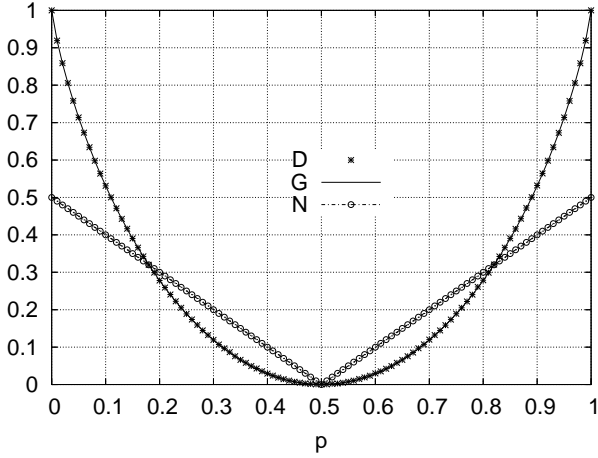}}
\caption{\label{fignume2}Plots of $D(\rho_\mathrm{b})$,
$G(\rho_\mathrm{b})$, and
$\mathcal{N}(\rho_\mathrm{b})$ against $p$. As we see in the text,
$D(\rho_\mathrm{b})=G(\rho_\mathrm{b})$ in this example.} 
\end{center}
\end{figure}

Figure \ref{fignume2} shows the plots of
$D(\rho_\mathrm{b})$, $G(\rho_\mathrm{b})$, and
$\mathcal{N}(\rho_\mathrm{b})$ against $p$ ($0\le p\le1$).
All of these measures vanish at $p=0.5$ and have positive values for
$p\not = 0.5$. Thus the difference among measures is not significant for
$\rho_\mathrm{b}$.

\begin{figure}[bpt]
\begin{center}
\scalebox{0.7}{\includegraphics{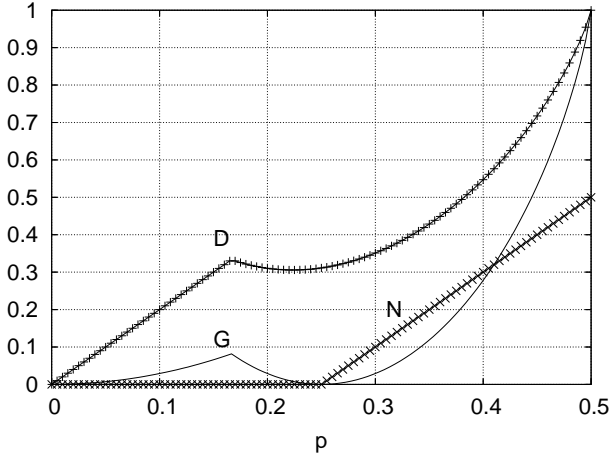}}
\caption{\label{fignume3}Plots of $D(\sigma)$, $G(\sigma)$, and
$\mathcal{N}(\sigma)$ against $p$.}
\end{center}
\end{figure}

The third example is the density matrix (this is also of a
$2\times 2$ system):
\begin{equation}
 \sigma=\left(\begin{array}{cccc}1/2-p&0&0&0\\
0&p&p&0\\
0&p&p&0\\
0&0&0&1/2-p
\end{array}\right)
\end{equation}
with $0\le p \le 1/2$. 
This may be seen as a mixture of pure states
$|00\rangle$, $|11\rangle$, and $(|01\rangle+|10\rangle)/\sqrt{2}$
with certain weights. It is separable for $p\le 1/4$ because
the eigenvalues of $(I\otimes\Lambda_\mathrm{T})\sigma$ are
$1/2-2p$, $p$ (with the multiplicity two), and $1/2$.
$D(\sigma)$ is, in contrast, non-zero unless $p=0$ as shown in
Fig.\ \ref{fignume3}.
In the figure, we also plot
\[
 G(\sigma)=\mathrm{min}\left\{1-H\left(\frac{1}{2}+p\right),
1-H(2p)\right\}.
\]
This is derived from the following values: the eigenvalues of
$\sigma$ are $0$, $1/2-p$ (with the multiplicity two), and $2p$;
the eigenvalue of $\mathrm{Tr}_B\sigma$ and that of $\mathrm{Tr}_A\sigma$
are both $1/2$ with the multiplicity two.
Interestingly, the shape of the curve of $G(\sigma)$ is similar to
that of $D(\sigma)$. One drawback in using $G(\sigma)$ is that it
vanishes at $p=0.25$ although $\sigma$ has no product eigenbasis for
this value.

\begin{figure}[bpt]
\begin{center}
\scalebox{0.7}{\includegraphics{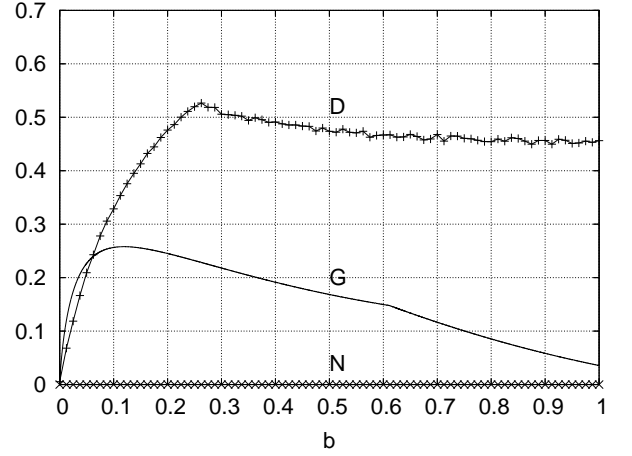}}\caption{\label{fignume4}
Plots of $D(\sigma_b^{[A,B]})$, $G(\sigma_b^{[A,B]})$, and
$\mathcal{N}(\sigma_b^{[A,B]})$ against $b$. Note that
$\mathcal{N}(\sigma_b^{[A,B]})$ is zero.}
\end{center}
\end{figure}

As the final example for the bipartite case, we consider the density
matrix of a $2\times 4$ system originally introduced by Horodecki \cite{H97}:
\begin{equation}\label{eqsigma}
 \sigma_b^{[A,B]}=\frac{1}{7b+1}\left(\begin{array}{cccccccc}
b&0&0&0&0&b&0&0\\
0&b&0&0&0&0&b&0\\
0&0&b&0&0&0&0&b\\
0&0&0&b&0&0&0&0\\
0&0&0&0&\frac{1+b}{2}&0&0&\frac{\sqrt{1-b^2}}{2}\\
b&0&0&0&0&b&0&0\\
0&b&0&0&0&0&b&0\\
0&0&b&0&\frac{\sqrt{1-b^2}}{2}&0&0&\frac{1+b}{2}
\end{array}\right)
\end{equation}
($0\le b\le 1$). This is known to be bound entangled for $0<b<1$, {\em
i.e.}, it is positive after partial transposition although it is
inseparable for $0<b<1$.
Hence $\mathcal{N}(\sigma_b^{[A,B]})=0$. We find, in contrast, that
$D(\sigma_b^{[A,B]})$ and $G(\sigma_b^{[A,B]})$ are non-zero for $b>0$
as shown in Fig.\ \ref{fignume4}.
The measure $G(\sigma_b^{[A,B]})$ is also plotted in the figure.
This is accomplished by using the definition Eq.\ (\ref{eqG}) together with the
following values. We have the eigenvalues of $\sigma_b^{[A,B]}$:
\[\begin{split}
 &\frac{2b+1\pm\sqrt{2b^2-2b+1}}{14b+2},~
\frac{b}{7b+1},~\frac{2b}{7b+1}~\mbox{(multiplicity two)},\\
&~\mathrm{and}~0~\mbox{(multiplicity three)};
\end{split}\]
the eigenvalues of $\mathrm{Tr}_B\sigma_b^{[A,B]}$:
$(3b+1)/(7b+1)$ and $4b/(7b+1)$; and the eigenvalues of
$\mathrm{Tr}_A\sigma_b^{[A,B]}$: $(3b+1\pm\sqrt{1-b^2})/(14b+2)$ and
$2b/(7b+1)$ with the multiplicity two.

\subsection{Tripartite examples}
Numerical computation of $D$ and $G$ defined by Eqs.\ (\ref{eqDM}) and
(\ref{eqGM}), respectively, is easy also for a tripartite density matrix
$\rho^{[1,2,3]}$ of three qubits.
We will compare $D(\rho^{[1,2,3]})$, $G(\rho^{[1,2,3]})$, and the
minimum and maximum negativities of $\rho^{[1,2,3]}$ over all
bipartite splittings.

Consider the pseudo-GHZ (PGHZ) state:
\[
\rho_\mathrm{PGHZ}=p|\psi_\mathrm{GHZ}\rangle\langle\psi_\mathrm{GHZ}| +
(1-p)\mathbbm{1}/8
\]
with $|\psi_\mathrm{GHZ}\rangle = (|000\rangle+|111\rangle)/\sqrt{2}$,
the Greenberger-Horne-Zeilinger (GHZ) state.
For this state, it is possible to find $G(\rho_\mathrm{PGHZ})$
analytically: The eigenvalues of $\rho_\mathrm{PGHZ}$ are
$(1+7p)/8$ and $(1-p)/8$ with the multiplicity seven for the latter one;
the reduced density matrix of any subpart has the eigenvalue $1/2$ with
the multiplicity two. Thus $G(\rho_\mathrm{PGHZ})=1-H[(1+p)/2]$. We can
easily find that this equation holds for any number of qubits
(larger than two) for the PGHZ state.

\begin{figure}[bpt]
\begin{center}
\scalebox{0.7}{\includegraphics{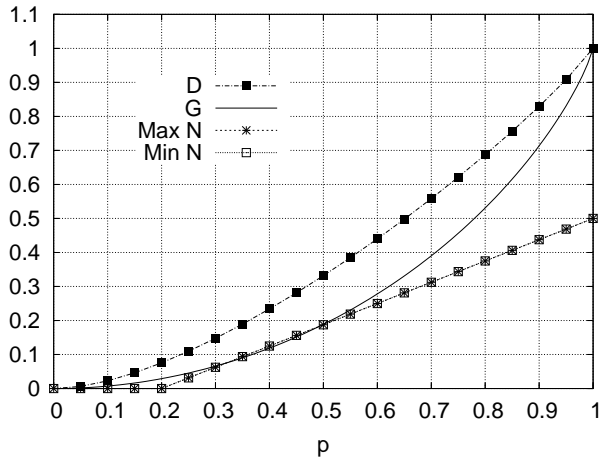}}
\caption{\label{fignume6}Plots of $D(\rho_\mathrm{PGHZ})$ and
$G(\rho_\mathrm{PGHZ})$ in comparison to
those of the minimum and the maximum of the negativity for $\rho_\mathrm{PGHZ}$
over all bipartite splittings (as functions of $p$).} 
\end{center}
\end{figure}

Figure \ref{fignume6} shows the plots of $D(\rho_\mathrm{PGHZ})$,
$G(\rho_\mathrm{PGHZ})$, and the minimum and maximum of the negativity
for $\rho_\mathrm{PGHZ}$ over all bipartite splittings against $p$
($0\le p\le1$).
$D$ and $G$ reflect nonclassical correlation for $\forall p$ except
for $p=0$.

\begin{figure}[bpt]
\begin{center}
\scalebox{0.7}{\includegraphics{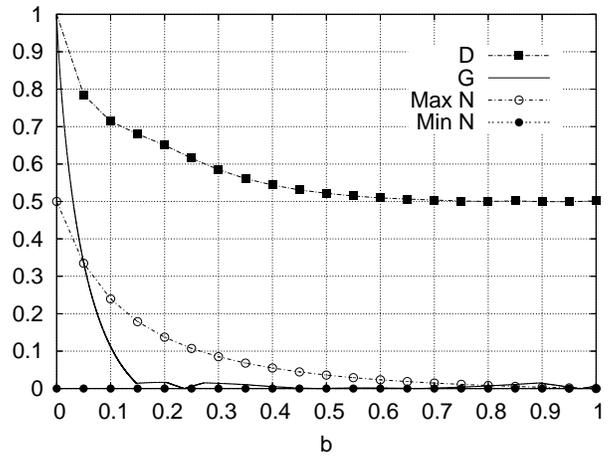}}\caption{\label{fignume5}
Plots of $D(\sigma_b^{[1,2,3]})$ and $G(\sigma_b^{[1,2,3]})$
in comparison to those of the minimum and the maximum of the negativity
for $\sigma_b^{[1,2,3]}$ over all bipartite splittings
(as functions of $b$).}
\end{center}
\end{figure}

The density matrix of the final example for a $2\times 4$ bipartite case
can be also interpreted as a density matrix of a $2\times 2\times 2$
tripartite case.
Let us consider the density matrix $\sigma_b^{[1,2,3]}$ of three qubits
given by the matrix of Eq.\ (\ref{eqsigma}). Figure \ref{fignume5} shows
the plot of $D(\sigma_b^{[1,2,3]})$, the plot of
$G(\sigma_b^{[1,2,3]})$, and the plots of the maximum and
minimum negativities over all bipartite splittings.
The computation of $G(\sigma_b^{[1,2,3]})$ is performed by using the
eigenvalues of $\sigma_b^{[1,2,3]}$ (found in the previous subsection) and 
the sets of eigenvalues of the reduced density matrices of subsystems $k$ 
($k=1,2,3$ and the following sets are given in this order):
$\{(4b)/(7b+1), (3b+1)/(7b+1)\}$, $\{1/2, 1/2\}$, and
$\{1/2, 1/2\}$.
It is found that $D(\sigma_b^{[1,2,3]})$ behaves similarly to the
maximum negativity over all bipartite splittings although the
convergence values are different.
The use of $G(\sigma_b^{[1,2,3]})$ seems to be improper for this example
because its value is very small for $b\gtrsim 0.15$ despite the fact
that $\sigma_b^{[1,2,3]}$ has no (fully) product eigenbasis for any $b$.
\section{Discussion}\label{sec5}
We have considered a nonclassical correlation based on a commonly
recognized paradigm other than the separability paradigm. We have
introduced two measures of nonclassical correlation for the paradigm
claiming that a multipartite system described by a density
matrix having no product eigenbasis possesses a nonclassical correlation.

A motivation to consider a paradigm other than the separability paradigm is
connected to a conceptual investigation of the separability paradigm.
Let us consider a bipartite system to simplify the discussion.
The separability paradigm says that a system described by a bipartite
separable density matrix can be prepared remotely when two distant
persons (Alice and Bob) receive instructions from a common source. Hence
there is no quantumness possessed by the system described by such a
density matrix in the context of remote preparation.
This assumes that a density matrix is a temporal average of
instantaneous density matrices or an ensemble average whose component
density matrices are accessible independently because Alice and Bob
prepare component states one by one.

Nevertheless, this seems to be a non-mixing process because
Alice and Bob have accesses to individual instances or components; this
is in contrast to usual processes in ensemble dynamics which are mixing.
To make it a mixing process, they should lose their memories about time
ordering of instances when the averaging is temporal averaging; in the
case of ensemble averaging, they should lose their memories about
indices of components.
In contrast, Alice and Bob do not have any memory to lose
when subparts of a system are distributed to them after a joint
preparation of a quantum state. Thus, it is questionable to compare 
the amount and quality of correlation of a remotely-prepared state with
those of a jointly-prepared state because the two contexts are different.

The paradigm that we support in this work assumes that a density
matrix having no product eigenbasis possesses nonclassical
correlation. A discussion on the state preparation is not involved in
its context. It is based on the problem as to whether or not
off-diagonal elements of the density matrix of a multipartite system can
be completely eliminated by local unitary operations.
This paradigm has been widely recognized in the community
while it is not considered to be a replacement of the separability
paradigm. It is a highly conceptual problem as to which protocol should
be a base protocol to think about correlation.

We have studied two measures $D$ and $G$ of non classical correlation
based on the paradigm. The former is defined as a global uncertainty
while the latter is defined as a local uncertainty. These two
measures are quite different in their naturalness: the former can be
interpreted as the minimum uncertainty for a global observer on the
total system after receiving reports on local systems; the latter
is based on some quite artificial game and is, of course, unnatural. It
is thus unexpected that plotted curves of these measures are sometimes
similar to each other as we have seen in numerical results.
We may enjoy the advantage to use the measure $G$, namely its easiness
of computation. In contrast to $D$, which requires a numerical search for
computing its value, $G$ can be computed by considering a finite number
of eigenvalue partitionings. A drawback is that $G$ is possibly small
for a density matrix having many non-zero eigenvalues, as we have seen
in Fig.\ \ref{fignume5} for an example. It is easy to produce mimic
eigenvalues of a subsystem close to genuine ones if the number of
possible partitionings is large. This drawback is related to the fact
that $G$ is not (fully or weakly) additive but subadditive, in
contrast to $D$ that is fully additive. A measure with (full or weak)
additivity is more reliable to quantify nonclassical correlation as
the system size grows.

As we mentioned, there is a similarity between $D$ and the measure
defined by Groisman {\em et al}. \cite{GKM07}. Let us consider a
bipartite case. Suppose that we change the definition of $D$ in the way
that we choose the basis written as a product of two eigenbases of local
systems $A$ and $B$ instead of searching the minimum over all product
bases. Then, this redefined measure is included in the distance
measures using the discrepancy between the given bipartite density matrix
$\rho^{[A,B]}$ and the Schmidt state $\rho_\mathrm{Sch}^{[A,B]}$. The
measure $D$ is thus close to the measures defined as a discrepancy
between $\rho^{[A,B]}$ and some specific density matrix having product
eigenbasis (this can be the one that minimizes the distance, or some
particular one to simplify the computing process).

Finally, we take a brief look at an ongoing development in
measures of nonclassical correlation. One of the measures very recently
proposed by Piani {\em et al.} \cite{PHH08} is designed to quantify
nonclassical correlation in the same paradigm as presently employed.
It is their measure $\Delta_{\rm CC}$ defined in a
similar way as that of quantum discord \cite{Z02}: $\Delta_{\rm CC}$
is a discrepancy between the quantum mutual information \cite{T02},
$I(\rho^{[A,B]})$, calculated for a bipartite density matrix
$\rho^{[A,B]}$ and $\text{min}_{\mathcal{M}_A, \mathcal{N}_B}I[
(\mathcal{M}_A\otimes\mathcal{N}_B)\rho^{[A,B]}]$ with two measurement
maps $\mathcal{M}_A$ and $\mathcal{N}_B$ associated to POVMs.
They also showed a straightforward extension of their theory to the
multipartite case. An advantage of $\Delta_{\rm CC}$ is that it vanishes
if and only if $\rho^{[A,B]}$ has a biproduct eigenbasis. A disadvantage
is the difficulty of finding the minimum as is similar to the above-discussed
disadvantage of $D$.

There must be a large number of measures to quantify nonclassical
correlation for the paradigm; this is reminiscent of the dawn of
entanglement measures. It is hoped that the paradigm will be studied
extensively to extend another branch of quantum information
science than the branch of the separability paradigm.

\section{Summary}\label{sec6}
Two measures of nonclassical correlation have been introduced to support
the paradigm claiming that a multipartite system described by a density matrix
having no product eigenbasis possesses nonclassical correlation.
The measure $D$ has been defined as the minimum uncertainty about a
joint system after we collect outcomes of particular local measurements.
The measure $G$ has been defined in the following way: consider the
minimum distance between a set of mimic eigenvalues and a set of genuine
eigenvalues of a local system on the basis of an artificial game. The
measure is defined by taking the maximum of this minimum distance over
all local systems. We have shown that $D$ is fully additive and $G$ is
subadditive.
Numerical computations of $D$ and $G$ have been performed by using a
random search of local bases and a non-probabilistic search of mimic
eigenvalues, respectively.
\begin{acknowledgments}
A.S. is thankful to Masahito Hayashi for a helpful comment on the
operational interpretation of one of the measures.
A.S. is supported by a Grant-in-Aid for JSPS Fellows
(Grant No. 1808962).
R.R. is supported by a Sasakawa Scientific Research Grant from JSS
and a Grant-in-Aid for JSPS Fellows (Grant No. 1907329).
M.N. would like to thank for partial supports of Grant-in-Aids
for Scientific Research from MEXT (Grant No. 13135215) and from
JSPS (Grant No. 19540422).
\end{acknowledgments}

\end{document}